\documentclass[12pt]{article}

\usepackage[numbers,sort&compress]{natbib}
\usepackage{graphicx}
\usepackage{amsmath}
\usepackage{amssymb}
\usepackage{hyperref}
\usepackage{bbm}

\ifx\hypersetup\sadfkjashdfkxja\else
\hypersetup{pdftitle={Deconstructing Supersymmetric S-matrices in D <= 2+1}}
\hypersetup{pdfauthor={Abhishek Agarwal and Donovan Young}}
\hypersetup{pdfsubject={}}
\hypersetup{pdfkeywords={}}
\hypersetup{plainpages=false}
\hypersetup{bookmarksnumbered=true}
\hypersetup{pdfstartview=FitH}
\hypersetup{pdfpagemode=UseNone}
\hypersetup{colorlinks=false}
\hypersetup{citebordercolor={.5 1 .5}}
\hypersetup{urlbordercolor={.5 1 1}}
\hypersetup{linkbordercolor={1 .7 .7}}
\fi

\def\be{\begin{equation}}
\def\ee{\end{equation}}
\def\bsp{\be\begin{split}}

\def\ra{\rangle}

\def\a{\alpha}
\def\b{\beta}
\def\g{\gamma}

\def\d{\delta}
\def\e{\epsilon}
\def\m{\mu}

\def\n{\nu}
\def\s{\sigma}
\def\r{\rho}
\def\l{\lambda}

\makeatletter

\newcommand{\Rmnum}[1]{\expandafter\@slowromancap\romannumeral #1@}
\makeatother

\newcommand{\beq}{\begin{equation}}
\newcommand{\eeq}{\end{equation}}
\newcommand{\bea}{\begin{eqnarray}}
\newcommand{\eea}{\end{eqnarray}}

\newcommand{\tprods}[2]{\langle#1#2\rangle}

\renewcommand{\title}[1]{\vbox{\center\LARGE{#1}}\vspace{5mm}}
\renewcommand{\author}[1]{\vbox{\center\large{#1}}\vspace{5mm}}
\newcommand{\address}[1]{\vbox{\center\em#1}}
\newcommand{\email}[1]{\vbox{\center\tt#1}\vspace{5mm}}

\newcommand{\matr}[2]{\left(\begin{array}{#1}#2\end{array}\right)}

\newcommand{\Tr}{\mathrm{Tr}}

\newcommand{\Cset}{{\,\,{{{^{_{\pmb{\mid}}}}\kern-.47em{\mathrm C}}}}}

\newcommand{\half}{\frac12}

\newcommand{\comment}[1]{}
\begin{document}
\newpage
\setcounter{page}{1}
\pagenumbering{arabic}
\renewcommand{\thefootnote}{\arabic{footnote}}
\setcounter{footnote}{0}

\begin{titlepage}
\title{\vspace{1.0in} {\bf Deconstructing Supersymmetric $S$-matrices in $D \leq 2+1$}}
 
\author{Abhishek Agarwal$^a$ and Donovan Young$^b$}

\address{$^a$Physical Review Letters, American Physical Society, 1
  Research Road, Ridge, NY 11961, USA\\ and\\ Physics Department, City
  College of CUNY, New York, NY 10031 USA\\
\vspace{.3cm} $^b$Niels Bohr Institute, University of Copenhagen,
Blegdamsvej 17, DK-2100 Copenhagen, Denmark}

\email{$^a$abhishek@ridge.aps.org, $^b$dyoung@nbi.dk}

\abstract{Global supersymmetries of the $S$-matrices of $\mathcal{N} =
  2,4,8$ supersymmetric Yang-Mills theories in three spacetime
  dimensions (without matter hypermultiplets) are shown to be $SU(1|1)
  $, $SU(2|2)$ and $SU(2|2) \otimes SU(2|2)$ respectively. These
  symmetries are not manifest in the off-shell Lagrangian formulations
  of these theories. A direct map between these symmetries and their
  representations in terms of the Yang-Mills degrees of freedom and
  the corresponding quantities in Chern-Simons-Matter theories with
  $\mathcal{N} \geq 4$ supersymmetry is also obtained. Dimensional
  reduction of the on-shell observables of the Yang-Mills theories to
  two spacetime dimensions is also discussed.}

\end{titlepage}

\section{Introduction and Summary}

In this paper we continue with investigations of hidden symmetries of
$S$-matrices of three dimensional supersymmetric Yang-Mills (SYM)
theories and their relation to the corresponding quantities for
supersymmetric Chern-Simons matter (SCS) theories.  In a previous
publication \cite{so(n)}, we showed that the $S$-matrices of SYM
theories with $\mathcal{N}\geq 2$ supersymmetry (without additional
matter hypermultiplets) have additional bosonic symmetries that are
not manifest in their off-shell Lagrangian formulations. In
particular, the bosonic symmetries of the $\mathcal{N} \geq 2$
$S$-matrices was shown to be $SO(\mathcal{N})$, while only a global
$SO(\mathcal{N}-1)$ $R$-symmetry is explicitly realized in the
Lagrangians. In this note we uncover the supersymmetric completion of
the bosonic symmetries of the $S$-matrices and find them to be
$SU(1|1)$, $SU(2|2)$, and $SU(2|2)\otimes SU(2|2)$ for $\mathcal{N} =2,4$
and $8$ SYM theories respectively.

A related class of gauge theories of much recent interest are SCS
models with $\mathcal{N} \geq 4$ supersymmetry. In particular the
$\mathcal{N}=6$ ABJM model \cite{abjm} and the $\mathcal{N} = 8$ BLG
theories \cite{blg} have been investigated in great detail in the
recent literature focusing on M2-branes.  For instance, the
$S$-matrix of the superconformal ABJM model has been shown to have
numerous fascinating hidden structures, including a potential infinite
dimensional Yangian symmetry \cite{SCS, scs-others}. Since the SYM and
SCS theories are expected to be related by renormalization group flows
(at least in the case of maximal supersymmetry through the flow of the
D2-brane theory to M2 ) one might expect some aspects of the
symmetries of the SCS $S$-matrices to be evident in on-shell
properties of the SYM theories as well. A puzzling aspect of the D2
to M2 flow is the lack of a direct off-shell connection between the
symmetries and degrees of freedom of the respective worldvolume
theories\footnote{The M2-brane theory can be shown to be related to
  the D2-brane theory through a Higgs mechanism \cite{m2d2}.}.  For
instance, the ABJM model has four complex scalars and a $SU(2|2)
\times g_2$ supersymmetry invariance, while the $\mathcal{N}=8$ SYM
theory on the other hand has a $SO(7)$ $R$-symmetry relating the seven
real scalars of the theory. Part of what we do in this paper is show
that the on-shell supersymmetry algebras of the SYM and SCS theories
can be mapped to one another. We also provide a dictionary
connecting the on-shell physical gauge invariant degrees of freedom of
these two classes of theories.

The organization of the paper and a summary of our results are as
follows. Starting with a particularly generic off-shell formulation of
the $\mathcal{N}=2,4,8$ SYM theories in three dimensions we review the
arguments from \cite{so(n)} showing that the $S$-matrices for these
theories possess a $SO(\mathcal{N})$ symmetry. We also briefly comment
upon the three dimensional analog of the spinor-``helicity'' formalism
developed in \cite{so(n)} that allows this symmetry enhancement to be
manifest.

In the next sub-section we briefly review some salient aspects of the
on-shell supersymmetry symmetry algebra of SCS theories with
$\mathcal{N} = 4,6$, and 8. In particular we focus on the $SU(2|2)$
structures that are naturally present as the symmetries of the
$S$-matrices of these theories. Much of our discussion on SCS theories
is based on the formalism introduced in \cite{SCS}.

The following section contains the central results of this note where
we present a clear connection between the the degrees of freedom and
the underlying supersymmetry algebras of the SYM and SCS theories. In
particular we show how to construct the $SO(\mathcal{N})$ covariant
``$\r$'' tensors -- that fix both the off-shell Lagrangians as well as
the on-shell SUSY algebra of the SYM theories -- from the single
particle representation of the on-shell SUSY algebra of the SCS
models.  As a result we are able to extend the results of \cite{so(n)}
and uncover the full global supersymmetry algebras for the SYM
theories which are shown to be $SU(1|1)$, $SU(2|2)$, and $SU(2|2)\otimes
SU(2|2)$ for $\mathcal{N} =2,4$, and 8 SYM theories
respectively\footnote{For other recent applications of $SU(2|2)$-type
  symmetries to studies of lower dimensional gauge theories
  see \cite{su22-other}.}. This is to be contrasted with the global
off-shell bosonic symmetries of these theories which are
$SO(\mathcal{N}-1)$. As noted in \cite{so(n)}, this symmetry
enhancement is due to $\mathcal{N}-1$ extra $U(1)$ generators that
couple the scalar degree of freedom arising from the on-shell gluon to
the scalars transforming under the $R$-symmetry generators. The
$\mathcal{N}-1$ extra $U(1)$ generators enhance the bosonic symmetries
to $SO(\mathcal{N})$ and the $SU(1|1), SU(2|2)$, and $SU(2|2)\otimes
SU(2|2)$ superalgebras are the supersymmetric completion of the
enhanced bosonic symmetries of the $S$-matrices. As a consequence of
this construction we are also able to relate the degrees of freedom of
the SYM and SCS theories; namely identify the degrees of freedom which
furnish a representation of the part of the superalgebra that is
common to both these theories. In the case of $\mathcal{N} = 4$
supersymmetry we find the complex combination of the real degrees of
freedom of the SYM theory that carry a representation of the $SU(2|2)$
superalgebra carried by the matter hypermultiplets of the SCS
theory. For $\mathcal{N} = 8$ supersymmetry, the same construction is
doubled, in a precise sense outlined later. Furthermore, we also find
the precise truncation, in terms of the SYM supercharges, of the
$\mathcal{N} = 8$ superalgebra to $\mathcal{N} = 6$ -- the
superalgebra of the ABJM theory.

In the final section of the paper we show that the on-shell
supersymmetry algebras of the SYM theories considered in this paper
survive a dimensional reduction to $1+1$ dimensions. In particular we
show how the dimensional reduction of the spinor-``helicity''
formalism allows one to eliminate the two dimensional gluon via gauge
transformations while turning the three-dimensional gluon into an
on-shell pseudo-scalar in two spacetime dimensions. We hope that this
note will be useful in the further analysis of the on-shell symmetries and
potential integrable structures for D2 and D1-brane theories.

\section{Invariant on and off shell formulations of $D = 2+1$ SYM and
  SCS theories}

We begin with a unified off-shell presentation of the $\mathcal{N} = 2,4,8$
SYM theories obtained in \cite{so(n)}. The action for these theories can be
written in the following compact notation ($a$ is a gauge-group index)
\bsp S &= -\frac{1}{4}\int
F_{\m\n}^aF^{a\m\n} -\frac{1}{2} \int
(D_\m\phi^i)^a(D^\m\phi^i)^a + \frac{i}{2} \int
\bar{\l}_I^a\g^\m D_\m\l^a_I \\& +\frac{i}{2}\int f^{abc}\r^i_{AB}\bar
\l ^a_A(\phi^i)^b\l^c_b - \frac{1}{4}\int f^{abc}f^{amn}(\phi
^i)^b(\phi ^j)^c(\phi ^i)^m(\phi ^j)^n.
\end{split}\label{3daction}
\ee
The capital indices $A,B,I$ run from $1\cdots \mathcal{N}$ while the
number of adjoint scalars (or the range of the lower case indices,
$i,j$) is $\mathcal{N}-1$. The global $R$ symmetries of these
theories are obviously $SO(\mathcal{N}-1)$.

While most of the terms in the action are the standard ones for
supersymmetric Yang-Mills theories in any number of dimensions, the
$\r$ tensors appearing as the Yukawa coupling are specific to three
dimensions. Their explicit form depends on how copies of three
dimensional gamma matrices are embedded in the gamma matrices of the
higher dimensional minimally supersymmetric theories, of which the
$D=3$ theories can be thought of as dimensional reductions.

As shown in \cite{so(n)}, the $\r$ tensors are key to understanding
how the on-shell $S$-matrices of these theories have an enhanced
symmetry, namely $SO(\mathcal{N})$. The hidden $SO(\mathcal{N})$
symmetry can readily be glimpsed by the following
observation. Combining the $\rho$ tensors that dictate the Yukawa
couplings with the obvious $SO(\mathcal{N})$ invariant, namely, the
delta function $\d_{AB}$, we get a tensor that has natural transformation
properties under $SO(\mathcal{N})$
\be 
\rho ^C_{AB} = \{\rho^1_{AB} = \delta_{AB}, \rho^i_{AB}\} \to
\r^D_{AC} \r^E_{BC} + \r^E_{AC} \r^D_{BC} = 2\d^{DE}\d_{AB} .
\ee 
For example, for $\mathcal{N}=2$ we have
\be  
\rho^C_{AB} = \{\delta_{AB}, \epsilon_{AB}\}, 
\ee
i.e. the two $SO(2)$ invariants. For $\mathcal{N}=8$, $\rho ^A_{BC}$
are the well known ${\bf 8}_{s,c,v}$ symbols relating the three eight
dimensional representations of $SO(8)$. As we shall see later on, the
$\r$ tensors also dictate the bosonic part of the on-shell symmetries
of these theories. A main result reported later in this paper is the
full supersymmetry algebra whose bosonic part -- namely $SO(\mathcal{N})$ --
is captured by the $\r$ tensors.

Before moving on to other issues we note that (\ref{3daction}) is
invariant under the following off-shell supersymmetry transformations
\bsp
&\d A^a_\m = -2i \bar \l _A\g_\mu \e_A, \\
&\d (\phi^i)^a = -2i\r^i_{AB}\bar\l _A^a\e_B,\\
& \d \l^a_A = F_{\m\n}^a\g^{\mu \n}\e_A + 2(D_\m \phi^i)^a\r^i_{AB}\g^\m \e_B - f^{abc}\r^i_{AB}\r^j_{BC} (\phi^i)^b(\phi^j)^c\e_C .
\end{split}\label{symsusy}
\ee
Supersymmetry invariance requires that the $\r$ tensors satisfy the
following identities\footnote{The signs in front of the epsilon
  tensors are sensitive to the ordering of the $\r^A$'s. We have used
  the conventions of section \ref{su22} below.}
\bsp\label{rt1}
&\r^i_{AB} = - \r^i_{BA},\\
& \r^i_{AB}\r^j_{CB} + \r^j_{AB}\r^i_{CB} = 2\d^{ij}\d_{AC},\\
&\r^{i}_{AB}\r^i_{CD} = \d_{AC}\d_{BD} - \d_{AD}\d_{BC} -\e_{ABCD},
\end{split}
\ee
for the case of $\mathcal{N} =2,4$. In the case of maximal
supersymmetry the last of the three identities needs to be modified to
\be\label{rt2}
\r^{i}_{AB}\r^i_{CD} = \d_{AC}\d_{BD} - \d_{AD}\d_{BC} 
-\e_{\hat A \hat B \hat C \hat D}
+\e_{\tilde A \tilde B \tilde C \tilde D}
-\e_{\hat A \hat B \tilde C \tilde D},
\ee
where the hatted indices run from $1,\ldots, 4$ and the tilded indices
from $5,\ldots,8$. The last term indicates $\e_{3456}$, and is totally
antisymmetric under permutations. Further details and additional
properties of the $\r$ tensors relevant to proving off-shell SUSY
invariance are provided in appendix \ref{app}.

Focusing now on on-shell quantities\footnote{For recent reviews of
  on-shell methods see \cite{onshell}.}, it is convenient to introduce a
three dimensional polarization vector
\be 
\epsilon_\mu (p,k) = \frac{\langle p|\gamma_\mu|k\rangle}{\tprods{k}{p}}, \hspace{.3cm}
p_\mu \epsilon^\mu (p,k) = k_\mu \epsilon^\mu (p,k) = 0, \label{pol}
\ee
(in the notation of \cite{so(n)}) for the oscillator expansion of the
gauge potential. $p$ is the physical momentum of the gluon while $k$
is an auxiliary momentum whose choice is tantamount to gauge
fixing. After carrying out the oscillator expansions, the
supersymmetry algebra (\ref{symsusy}) translates into the following
$SO(\mathcal{N})$ covariant transformations for the on-shell fields
\be 
Q_A|a^B\rangle = \frac{u}{2}\rho^B_{AC}|\l_C\rangle, \hspace{.3cm}
Q_A|\l_B\rangle = -\frac{u}{2}\rho^C_{AB}|a^C\rangle.\label{onshell}
\ee 
$|a^1\ra$ is the on-shell scalar obtained from the gluon, while
$|a^2\ra ,\ldots, |a^{\mathcal{N}}\ra$ are the on-shell versions of
the real scalars present in the Lagrangian\footnote{This symmetry can
  also be uncovered upon a linearization of the recently constructed
  gauge invariant formalism for SYM theories. For a discussion on this
  matter, we refer to \cite{AA-VPN}; especially the last section of
  this reference.}. $u$ is a real three dimensional Majorana spinor,
whose form in the conventions of \cite{so(n)} is
\be 
u(p) =
\frac{1}{\sqrt{p_0 - p_1}}\matr{c}{p_2 \\ p_1 - p_0}.\label{u} 
\ee
Since the supersymmetry algebra above is manifestly $SO(\mathcal{N})$
covariant, the $S$-matrix for these theories must necessarily be
$SO(\mathcal{N})$ invariant for it to commute with the on-shell
supercharges. This statement was also illustrated explicitly at the
level of the four particle amplitudes in \cite{so(n)}.

As is evident, the hidden enhanced bosonic symmetry of the SYM
$S$-matrices are encoded in the the $\r$ tensors, which also fix the
Yukawa couplings of the corresponding off-shell Lagrangians. In
section \ref{su22} we shall construct these tensors from
representations of the on-shell superalgebras relevant to SCS theories
which will allow us to both find the supersymmetric completion of the
bosonic $SO(\mathcal{N})$ symmetries and relate the superalgebras
underlying the $S$-matrices of the SYM and SCS theories.

\subsection{Supersymmetric Chern-Simons-Matter theories}

Before relating the symmetries of the $S$-matrices of SYM and SCS
theories in the next section, let us briefly review some details of
Chern-Simons matter theories with $\mathcal{N} \geq 4$ supersymmetry.
In the case of $\mathcal{N} =4$ supersymmetry, one has two complex
scalars $\phi_a$ and two compensating fermionic degrees of freedom
$\psi _{\dot a}$ transforming under two different $SU(2)$ groups
(denoted by the dotted and undotted indices) \cite{gw}.  Since we
shall be concerned only with color ordered amplitudes, we will not
delve into the possible gauge groups and the representations
compatible with $\mathcal{N}\geq 4$ supersymmetry, except to refer to
\cite{SCS,gw,hlllp1,hlllp2}. One can add twisted matter
hypermultiplets $\tilde \phi _{\dot a}, $ $\tilde \psi_a$ which, in
general, can carry a different representation of the gauge group
without losing $\mathcal{N}=4$ supersymmetry
\cite{hlllp1,hlllp2}. However, when the twisted and untwisted
hypermultiplets carry the same representation, one has $\mathcal{N}=5$
supersymmetry. The special case of the hypermultiplets being in the
bifundamental representation of $SU(N)$ corresponds to the ABJM model
with $\mathcal{N}=6$ superconformal invariance \cite{abjm}, while the
particular case of $N = 2$ produces the maximally supersymmetric
$\mathcal{N} = 8$ BLG theory \cite{blg}. In the absence of twisted
hypermultiplets, one has four supercharges\footnote{$\a$ is the three
  dimensional Lorentz index.} $\mathcal{Q}_{\a b\dot{c}}$ which act
linearly on the on-shell fields as \cite{SCS}
\beq \label{abN4s}
\mathcal{Q}_{\a b\dot{c}}|\phi_d\rangle = \e_{bd}u_\a
|\psi_{\dot{c}}\rangle, \hspace{.3cm} \mathcal{Q}_{\a
  b\dot{c}}|\psi_{\dot{d}}\rangle =
\e_{\dot{c}\dot{d}}u_\a|\phi_b\rangle ,
\eeq
where $u_\a$ are solutions of the massless Dirac equation
in three dimensions (\ref{u}). The on-shell supersymmetry algebra for
the conformal $\mathcal{N} = 4$ SCS theories is simply $SU(2|2)$
\beq 
\{\mathcal{Q}_{\a b\dot{c}}, \mathcal{Q}_{\b e\dot{f}}\} =
\e_{be}\e_{\dot{c}\dot{f}}P_{\a\b} .
\eeq
In the case of $\mathcal{N} = 6$ supersymmetry one has two additional
supercharges $\tilde{\mathcal{Q}}_\a^\pm$ forming a $g_2$ algebra
which relate the twisted and untwisted matter fields. In the case of
maximal supersymmetry one has two complete copies of $SU(2|2)$
furnishing the eight supercharges needed for the BLG theory.  For a
detailed exposition of the on-shell symmetry algebra for a more
general class of theories (which include potential mass-deformations)
we refer to \cite{SCS}, where a Lagrangian formulation of these gauge
theories can also be found.

In the case of the SCS theories, the on-shell symmetries are also
reflected in the off-shell Lagrangians, which is in contrast to the
case of SYM theories \cite{so(n)}. In the next section, we show that
the underlying on-shell supersymmetry algebras of the $\mathcal{N} =
4$ and $\mathcal{N} = 8$ SYM and SCS theories are the
same. Furthermore, we obtain the appropriate truncation of the
$\mathcal{N}=8$ superalgebra of the SYM theories that reproduces the
$\mathcal{N} = 6$ on-shell algebra of the ABJM models. In the process,
we also obtain a precise map between the on-shell gauge invariant
degrees of freedom of these two different classes of gauge theories.

\section{$SU(2|2)$ structure for $S$-matrices}\label{su22}

In this section we will show that the on-shell superalgebra
(\ref{onshell}) respected by the S-matrices of the ${\cal N}=4$
(${\cal N}=8$) SYM theories may be re-cast into the $SU(2|2)$
($SU(2|2)\otimes SU(2|2)$) supersymmetry algebras obeyed by the ${\cal
  N}=4$ (${\cal N}=8$) SCS theories considered in \cite{SCS}. We will
also show that the intermediate case, ${\cal N}=6$, corresponds to
removing two of the 8 SUSY generators of the ${\cal N}=8$ SYM theory.

The on-shell degrees of freedom of the ${\cal N}=2,4,8$ SYM theory
consist of ${\cal N}$ real bosons $|a^A\ra$ and an equal number of
Majorana fermions $|\l_C\ra$. The superalgebra is then given by the
action of the supercharges $Q_{B,\a}$ upon these on-shell states
\cite{so(n)}
\be\label{SYMSUSY}
Q_{B,\a}|a^A\ra = \frac{1}{2}u_\a\,\r^A_{BC} |\l_C\ra,\qquad
Q_{B,\a}|\l_C\ra = -\frac{1}{2}u_\a\,\r^A_{BC}|a^A\ra ,
\ee
where the indices $A,B,C = 1,\ldots,{\cal N}$ and $u_\a$ is a 2-spinor
($\a=1,2$), while $\r^1=\mathbbm{1}$. The $\rho^I$ obey 
\be\label{rhorel}
\rho^I (\r^J)^T + \rho^J(\r^I)^T = 2 \d^{IJ} \mathbbm{1}.
\ee 

As discussed in the previous section, the SCS theories considered in
\cite{SCS} have, for the ${\cal N}=4$ theory, on-shell complex bosons
$|\phi_{b}\ra$ and complex fermions $|\psi_{\dot b}\ra$, obeying the
algebra (\ref{abN4s}). In the case of ${\cal N}=6$ and ${\cal N}=8$
there are also twisted versions of these states, denoted with a
tilde. In these cases the states also carry an additional label
corresponding to representations of the $g_2$ and $g_4$ algebras for
the ${\cal N}=6$ and ${\cal N}=8$ cases respectively \cite{SCS}. In
\cite{SCS} this additional label was denoted with a $\pm$ for the
${\cal N}=6$ case,
\be
|\phi_{b\pm}\ra, ~|\tilde\phi_{\dot b\pm}\ra,~|\psi_{\dot b \pm}\ra,~
|\tilde\psi_{b\pm}\ra,
\ee
and by the addition of an extra $SU(2)$ index (hatted for the
untwisted states and tilded for the twisted ones) for the ${\cal N}=8$ case 
\be
 |\phi_{b\hat c}\ra, ~|\tilde\phi_{\dot b\tilde c}\ra,~|\psi_{\dot b \hat c}\ra,~
|\tilde\psi_{b\tilde c}\ra.
\ee
It should be noted that the field content of the theory under
consideration is obtained by fixing a value for these extra
indices. For example, in the ${\cal N}=6$ case choosing $+$ or $-$
amounts to a choice of sign for the central charge in the algebra
\cite{SCS}. The ${\cal N}=4$ fields have no such extra indices, but it
is convenient for us to decorate them as in the ${\cal N}=8$ case,
with a (in this case superfluous) extra index, so that (\ref{abN4s}) becomes
\be\label{N4SCS}
{\cal Q}_{\a b\dot c} |\phi_{d \hat e}\ra = u_\a \e_{bd} |\psi_{\dot c
  \hat e} \ra,\quad
{\cal Q}_{\a b\dot c} |\psi_{\dot d \hat e}\ra = u_\a \e_{\dot c\dot d} |\phi_{b
  \hat e} \ra.
\ee
The ${\cal N}=8$ case essentially amounts to a doubling of this
algebra, and the addition of new supercharges $\tilde{\cal Q}_{\a
  \tilde b\hat c}$ which relate twisted and untwisted states. This
gives the $SU(2|2)\otimes SU(2|2)$ algebra
\bsp\label{SCSMSUSY}
&{\cal Q}_{\a b\dot c} |\phi_{d \hat e}\ra = u_\a \e_{bd} |\psi_{\dot c
  \hat e} \ra,\quad
{\cal Q}_{\a b\dot c} |\tilde \phi_{\dot d \tilde e}\ra = 
u_\a \e_{\dot c \dot d} |\tilde \psi_{b \tilde e} \ra,\\
&{\cal Q}_{\a b\dot c} |\psi_{\dot d \hat e}\ra = u_\a \e_{\dot c\dot d} |\phi_{b
  \hat e} \ra,\quad
{\cal Q}_{\a b\dot c} |\tilde \psi_{d \tilde e}\ra = 
u_\a \e_{bd} |\tilde \phi_{\dot c \tilde e} \ra,\\
&\tilde{\cal Q}_{\a \tilde b\hat c} |\phi_{d \hat e}\ra = u_\a
\e_{\hat c \hat e} |\tilde \psi_{d
  \tilde b} \ra,\quad
\tilde{\cal Q}_{\a \tilde b\hat c} |\psi_{\dot d \hat e}\ra = -u_\a
\e_{\hat c \hat e} |\tilde \phi_{\dot d
  \tilde b} \ra,\quad\\
&\tilde{\cal Q}_{\a \tilde b\hat c} |\tilde \psi_{d \tilde e}\ra = u_\a
\e_{\tilde b \tilde e} |\phi_{d\hat c} \ra,\quad
\tilde{\cal Q}_{\a \tilde b\hat c} |\tilde \phi_{\dot d \tilde e}\ra = -u_\a
\e_{\tilde b \tilde e} |\psi_{\dot d\hat c} \ra.
\end{split}
\ee

We now give an explicit map between the on-shell degrees of freedom of
the ${\cal N}=8$ SCS and SYM theories which translates the SUSY
algebras (\ref{SYMSUSY}) and (\ref{SCSMSUSY}) into one another; the
cases with less supersymmetry then follow in a straightforward way. We
begin by breaking-up the $SO(8)$ indices $A$, $B$, and $C$ in
(\ref{SYMSUSY}) into two indices, so that $A=(\hat A,\,\tilde
A)=(1,\ldots,4,\,5,\ldots,8)$, etc.. We then take two copies of the
the Pauli matrices $\s^i$ along with the unit matrix
\be
\s_{\hat A}=\s_{\tilde A} = (\mathbbm{1},~i\s^1,~i\s^2,~i\s^3).
\ee
Using the map\footnote{We suppress the spinor index $\a$ on the
  supercharges while the SU(2) indices of the SCS fields are
  understood to be carried by the Pauli matrices, the first index
  corresponding to the row. Furthermore, $\e\equiv i\s^2$ is
  understood to act on the Pauli matrices by usual matrix
  multiplication.}
\bsp
&{\cal Q} = Q_{\hat A} \, \s_{\hat A},\qquad
\tilde {\cal Q} = Q_{\tilde A} \, \s_{\tilde A},\qquad
|\phi\ra = |a^{\hat A}\ra \, \s_{\hat A},\qquad
|\tilde\phi\ra = |a^{\tilde A}\ra \, \s_{\tilde A},\\
&|\psi\ra = |\l_{\hat A}\ra\, \s_{\hat A}^T \,\e,\qquad
|\tilde\psi\ra = -|\l_{\tilde A}\ra \,\e\,\s_{\tilde A}^T,
\end{split}
\ee
one finds that
\bsp
&\r^{\hat A}_{\hat B \hat C} = -\frac{1}{2}\Tr \left(
\e \left[\s_{\hat C}^T\right]^{-1} 
\s_{\hat B}^T\, \e \,\s_{\hat A}\right),\\
&\r^{\hat A}_{\tilde B \tilde C} = -\frac{1}{2}\Tr \left(
\e \,\s_{\tilde B}^T
\left[\s_{\tilde C}^T\right]^{-1}  \e \,\s_{\hat A}\right),\\
&\r^{\tilde A}_{\hat B \tilde C} = \frac{1}{2}\Tr \left(
\left[\s_{\tilde C}^T\right]^{-1}\e\,
\s_{\hat B}\,  \e \,\s_{\tilde A}\right),\\
&\r^{\tilde A}_{\tilde B \hat C} = -\frac{1}{2}\Tr \left(
\s_{\tilde A}\,\e\,
\s_{\tilde B}\,
\e\,\left[\s_{\hat C}^T\right]^{-1}\right).
\end{split}
\ee
It is then straightforward to verify that $\r^1=\mathbbm{1}$ and that
(\ref{rt1}), (\ref{rt2}), and (\ref{SYMSUSY}) are obeyed. The map is given more
explicitly by
\bsp
&|a_1\ra = \frac{1}{2}(|\phi_{11}\ra+|\phi_{22}\ra),\quad
|a_2\ra = -\frac{i}{2}(|\phi_{12}\ra+|\phi_{21}\ra),\quad\\
&|a_3\ra = \frac{1}{2}(|\phi_{12}\ra-|\phi_{21}\ra),\quad
|a_4\ra = -\frac{i}{2}(|\phi_{11}\ra-|\phi_{22}\ra),\\
&|a_5\ra = \frac{1}{2}(|\tilde\phi_{11}\ra+|\tilde\phi_{22}\ra),\quad
|a_6\ra = -\frac{i}{2}(|\tilde\phi_{12}\ra+|\tilde\phi_{21}\ra),\quad\\
&|a_7\ra = \frac{1}{2}(|\tilde\phi_{12}\ra-|\tilde\phi_{21}\ra),\quad
|a_8\ra = -\frac{i}{2}(|\tilde\phi_{11}\ra-|\tilde\phi_{22}\ra),\\
\end{split}
\ee
\bsp
&Q_1 = \frac{1}{2}({\cal Q}_{11}+{\cal Q}_{22}),\quad
Q_2 = -\frac{i}{2}({\cal Q}_{12}+{\cal Q}_{21}),\quad\\
&Q_3 = \frac{1}{2}({\cal Q}_{12}-{\cal Q}_{21}),\quad
Q_4 = -\frac{i}{2}({\cal Q}_{11}-{\cal Q}_{22}),\\
&Q_5 = \frac{1}{2}(\tilde{\cal Q}_{11}+\tilde{\cal Q}_{22}),\quad
Q_6 = -\frac{i}{2}(\tilde{\cal Q}_{12}+\tilde{\cal Q}_{21}),\quad\\
&Q_7 = \frac{1}{2}(\tilde{\cal Q}_{12}-\tilde{\cal Q}_{21}),\quad
Q_8 = -\frac{i}{2}(\tilde{\cal Q}_{11}-\tilde{\cal Q}_{22}),
\end{split}
\ee
\bsp
&|\l_1\ra = \frac{1}{2}(|\psi_{12}\ra-|\psi_{21}\ra),\quad
|\l_2\ra = \frac{i}{2}(|\psi_{11}\ra-|\psi_{22}\ra),\quad\\
&|\l_3\ra = \frac{1}{2}(|\psi_{11}\ra+|\psi_{22}\ra),\quad
|\l_4\ra = -\frac{i}{2}(|\psi_{12}\ra+|\psi_{21}\ra),\\
&|\l_5\ra = -\frac{1}{2}(|\tilde\psi_{12}\ra-|\tilde\psi_{21}\ra),\quad
|\l_6\ra = \frac{i}{2}(|\tilde\psi_{11}\ra-|\tilde\psi_{22}\ra),\quad\\
&|\l_7\ra = -\frac{1}{2}(|\tilde\psi_{11}\ra+|\tilde\psi_{22}\ra),\quad
|\l_8\ra = -\frac{i}{2}(|\tilde\psi_{12}\ra+|\tilde\psi_{21}\ra).
\end{split}
\ee
The ${\cal N}=4$ case then follows simply by deleting the $\tilde
\phi$, $\tilde \psi$, and $\tilde Q$, or equivalently, by ignoring the
$\tilde A,\tilde B,\tilde C$ indices. The ${\cal N}=2$ case is a
further truncation of this, where in addition, the range of the
$\hat A,\hat B,\hat C$ indices is taken to be from 1 to 2, or
equivalently, where we take
\bsp
&{\cal Q}_{11}\to{\cal Q}_{22},\qquad
{\cal Q}_{12} \to {\cal Q}_{21},\qquad
|\phi_{11}\ra \to |\phi_{22}\ra,\qquad
|\phi_{12}\ra \to |\phi_{21}\ra,\\
&|\psi_{22}\ra \to -|\psi_{11}\ra,\qquad
 |\psi_{21}\ra \to -|\psi_{12}\ra.
\end{split}
\ee
Then the relations
(\ref{N4SCS}) reduce to a $SU(1|1)$ algebra.

The ${\cal N}=6$ SCS theory has a superalgebra given by \cite{SCS}
\bsp\label{N6SCS}
&{\cal Q}_{\a b\dot c} |\phi_{d \pm}\ra = u_\a \e_{bd} |\psi_{\dot c
  \pm} \ra,\quad
{\cal Q}_{\a b\dot c} |\tilde \phi_{\dot d \pm}\ra = 
u_\a \e_{\dot c \dot d} |\tilde \psi_{b \pm} \ra,\\
&{\cal Q}_{\a b\dot c} |\psi_{\dot d \pm}\ra = u_\a \e_{\dot c\dot d} |\phi_{b
  \pm} \ra,\quad
{\cal Q}_{\a b\dot c} |\tilde \psi_{d \pm}\ra = 
u_\a \e_{bd} |\tilde \phi_{\dot c \pm} \ra,\\
&\tilde{\cal Q}^\pm_\a |\phi_{b\mp}\ra = u_\a |\tilde\psi_{b\mp}\ra,
\qquad
\tilde{\cal Q}^\pm_\a |\psi_{\dot b\mp}\ra = -u_\a |\tilde\phi_{\dot b\mp}\ra,\\
&\tilde{\cal Q}^\mp_\a |\tilde\psi_{b\mp}\ra = -u_\a |\phi_{b\mp}\ra,
\qquad
\tilde{\cal Q}^\mp_\a |\tilde\phi_{\dot b\mp}\ra = u_\a |\psi_{\dot b\mp}\ra,\\
\end{split}
\ee
with all other actions of the supercharges upon the states producing
zero. We make the following map between the $\pm$ index on the scalars
and fermions and the tilded and hatted indices of (\ref{SCSMSUSY})
\be
+ \to 1,\qquad
- \to 2,
\ee
and then (\ref{N6SCS}) are equivalent to (\ref{SCSMSUSY}) with 
\be
\tilde{\cal Q}^+ \to \tilde {\cal Q}_{21},\qquad
\tilde{\cal Q}^- \to -\tilde {\cal Q}_{12},\qquad
\tilde{\cal Q}_{11} \to 0,\qquad \tilde{\cal Q}_{22}\to 0,
\ee
where the spinor index $\a$ has been suppressed. Therefore the
restriction to ${\cal N}=6$ is achieved by removing $\tilde{\cal
  Q}_{11}$ and $\tilde {\cal Q}_{22}$. In the SYM language we lose
$Q_5$ and $Q_8$ while
\be
\tilde {\cal Q}^+ = iQ_6 - Q_7,\qquad
\tilde {\cal Q}^- = -iQ_6 - Q_7.
\ee

The algebras mentioned in this section largely constrain the structure
of the $S$-matrices. In the case of four-particle amplitudes, the
supersymmetry algebra constrains the $S$-matrix to one undetermined
function of the coupling constant and the kinematic Mandelstam
variables.  In the notation of \cite{SCS}, one can decompose the
scattering matrix $S$ as $S = I + i\mathcal{T}$, where $S$ is the
scattering operator. The $SU(2|2)$ symmetry for the the $\mathcal{N} =
4$ SCS theory then allows one to parametrize the four-particle
amplitudes in terms of ten independent functions of the Mandelstam
variables $A,\ldots, L$ as
\bsp 
\langle \mathcal{T}|\phi_\a\phi_\b\phi_\g\phi_\d\rangle &=
\left[\half(A+B)\e_{\a\d}\e_{\b\g} +
  \half(A-B)\e_{\a\g}\e_{\b\d}\right]\d^3(\sum_ip_i),\\ \langle
\mathcal{T}|\psi_{\dot
  \a}\psi_{\dot\b}\psi_{\dot\g}\psi_{\dot\d}\rangle &=
\left[\half(D+E)\e_{\dot\a\dot\d}\e_{\dot\b\dot\g} +
  \half(D-E)\e_{\dot\a\dot\g}\e_{\dot\b\dot\d}\right]\d^3(\sum_ip_i),\\ \langle
\mathcal{T}|\phi_{\a}\psi_{\dot\b}\phi_{\g}\psi_{\dot\d}\rangle &=
-G\e_{\a\g}\e_{\dot \b\dot\d}\d^3(\sum_ip_i),\\ \langle
\mathcal{T}|\psi_{\dot \a}\phi_{\b}\psi_{\dot\g}\phi_{\d}\rangle &=
-L\e_{\dot\a\dot\g}\e_{\b\d}\d^3(\sum_ip_i),\\ \langle
\mathcal{T}|\phi_{ \a}\phi_{\b}\psi_{\dot\g}\psi_{\dot\d}\rangle &=
-\half C \e_{\a\b}\e_{\dot\g\dot\d}\d^3(\sum_ip_i),\\ \langle
\mathcal{T}|\phi_{ \a}\psi_{\dot\b}\psi_{\dot\g}\phi_{\d}\rangle &=
-H\e_{\a\d}\e_{\dot\b\dot\g}\d^3(\sum_ip_i),\\ \langle
\mathcal{T}|\psi_{\dot \a}\psi_{\dot\b}\phi_{\g}\phi_{\d}\rangle &=
-\half F\e_{\dot a\dot \b}\e_{\g\d}\d^3(\sum_ip_i),\\ \langle
\mathcal{T}|\psi_{\dot \a}\phi_{\b}\phi_{\g}\psi_{\dot\d}\rangle &=
-K\e_{\dot\a\dot\d}\e_{\b\g}\d^3(\sum_ip_i),
\end{split}
\ee
where we have suppressed the extra index corresponding to the
$g_{{\cal N}-4}$ representation, i.e. we have fixed the choice of $+$
or $-$ for the ${\cal N}=6$ case or fixed the hatted and tilded
indices to 1 or 2 in the ${\cal N}=8$ case. All the functions can be
expressed in terms of a single function (chosen to be $A$ in
\cite{SCS}) using the constraining properties of the algebra
alone. For instance, $D= -A\frac{\tprods{3}{4}}{\tprods{1}{2}}$, $G=
+A\,\frac{\tprods{ 4}{1}}{\tprods{1}{2}}$, etc. \cite{SCS}. Similar
constraints for $\mathcal{N} = 5,6$, and $8$ SCS theories were also
obtained in \cite{SCS}. On the other hand, amplitudes for SYM theories
with extended supersymmetry were constrained in a similar fashion
using a real basis for the scalar and fermion fields in
\cite{so(n)}. The map between the SYM and SCS on-shell degrees of
freedom obtained earlier in this section implies that these
constraints are common to both families of theories, as long as the
underlying on-shell superalgebras can be mapped to each other, namely,
in the cases of $\mathcal{N} = 4,6,$ and 8 supersymmetry.
\section{Dimensional reduction to $d=2$}

In this final section we note that the three dimensional on-shell
techniques can be easily reduced to two spacetime dimensions to make
the $SO(\mathcal{N})$ invariance of the lower dimensional gauge
theories manifest at the level of $S$ matrices.  We implement
dimensional reduction by compactifying the ``1'' direction. The real
three dimensional Majorana spinor (\ref{u}) becomes
\be 
\tilde u(p) = \frac{1}{\sqrt{p_0 }}\matr{c}{p_2 \\ - p_0} =
\sqrt{p_0} \matr{c}{sgn(p) \\ - 1} ,
\ee
upon dimensional reduction. The sign refers to the two dimensional
mass-shell condition $p_2 = sgn(p) p_0$. Under the action of the two
dimensional ``gamma-five'' (which is $\g^1$ in the three dimensional
notation), $\g^1\tilde{u} = -sgn(p)\tilde u$. Now, the $D=3$
polarization vector (\ref{pol}) can be simplified by choosing the
auxiliary momentum $k$ judiciously after reducing it to three
dimensions. Choosing $k_2 = p_0$ and $k_0 = -p_2$ makes $\e_0$ and
$\e_2$ vanish. Recalling that fixing $k$ is tantamount to choosing a
gauge, the above statement is nothing but an illustration of the fact
that the two dimensional gluon can be gauge transformed away. Under
the same gauge choice, $\e_1 = -sgn(p)$. Using the dimensionally
reduced versions of the three-dimensional quantities in the relation
$\delta A_1 = \delta \phi^1 = \frac{1}{2}(\bar \e_I \g_1 \l_I)$, and
employing the mode expansions given above, we get
\be Q_A |a^1\rangle
= \frac{\tilde u}{2}|\l_A\rangle ,
\ee 
in $D=2$. The dimensional reductions of the other mode expansions are
trivial, and they yield the $SO(\mathcal{N})$ covariant algebra
\be Q_A|a^B\rangle =
\frac{\tilde u}{2}\rho^B_{AC}|\l_C\rangle, \hspace{.3cm}
Q_A|\l_B\rangle = -\frac{\tilde
  u}{2}\rho^C_{AB}|a^C\rangle,\label{onshell2} 
\ee 
in two dimensions as expected. This reduction makes the
$SO(\mathcal{N})$ structure, and by the analysis presented earlier in
the paper, the $SU(1|1)$, $SU(2|2)$, and $SU(2|2)\otimes SU(2|2)$
symmetries of the $S$-matrices of the dimensional reductions of three
dimensional $\mathcal{N} = 2, 4,$ and $8$ SYM theories to two
spacetime dimensions manifest.

\section*{Acknowledgments}

DY was supported by FNU through grant number 272-08-0329.


\appendix

\section{Off-shell supersymmetry via $\r$ tensors}
\label{app}

In this section we provide further details of how supersymmetry
invariance of the action (\ref{3daction}) depends on the specific
properties of the $\r$ tensors (\ref{rt1})-(\ref{rt2}). While most of
the cancelations necessary to see the supersymmetry invariance of the
action are obtained using the properties of the $\r$ tensors mentioned
before, the cancelation of the variation of the Yukawa term imposes
some further constraints. The variation of the Yukawa term produces
\be
-\frac{i}{2}\int [\r^i\r^{[k}\r^{l]}]_{CD}f^{amn}f^{apg}
(\bar{\e}_D\l^n_C)(\phi^k)^p (\phi^l)^q(\phi^i)^m\label{yvar}.
\ee
The $\r$ tensors are multiplied in the expression above in the sense
of matrix multiplication, where for each value of $i$, $\r^i$ is an
$\mathcal{N} \times \mathcal{N}$ matrix.  For this term to cancel
against the variation of the $\phi^4$ term, we shall need to reduce
the term cubic in the $\r$ tensors to fewer factors of $\r$. To see
this in the case of $\mathcal{N}=4$ supersymmetry one needs to utilize
the fact that the tensors obey the $SO(3)$ algebra 
\be \r^i\r^j =
-\e^{ijk}\r^k, \hspace{.3cm} (\r^i)^2 = -I. 
\ee 
Using these identities, and carrying out the sum over the $i,k,l$
indices in (\ref{yvar}) produces two terms, $T_1$ and $T_2$. $T_1$
corresponds to the case where $i = k$ or $i=l$
\be T_1 = -2i\int
f^{abc}f^{amn}\r^i_{CD}(\bar \l _C\e_D)(\phi^j)^c(\phi^i)^m(\phi ^j)^n,
\ee
and it cancels against the variation of the $\phi^4$ term. $T_2$
corresponds to the case of $i\neq k \neq l$
\be T_2 = -i\int (\bar \l
_C\e_C)^n\left[f^{amn}f^{apl} + f^{apn}f^{alm} + f^{aln}f^{amp}
  \right](\phi^1)^p(\phi^2)^l(\phi^3)^m .
\ee 
The combination of structure constants in the square brackets vanishes
due to the Jacobi identity, hence 
\be T_2 = 0. 
\ee
To see how the ${\cal N}=4$ cancellation generalizes to the ${\cal N}=8$ case we
note the following properties. Let the indices $i$, $j$, and $k$ be
distinct. Then 
\be
\r^{ij} \equiv \frac{1}{2}\left(\r^i \r^j - \r^j\r^i\right) = \r^i\r^j
= -\r^j\r^i.
\ee
Consider
\be\label{rel}
\r^{ij} \, M = \r^k, \implies M = \left(\r^{ij}\right)^{-1}\r^k.
\ee
Since $(\r^i)^2=-\mathbbm{1}$ (no sum over $i$), we see that
$(\r^i)^{-1} = -\r^i = (\r^i)^T$. Therefore
\be
M = -\r^i\r^j\r^k.
\ee
We would now like to prove that the first relation in (\ref{rel}) is
cyclically invariant. So we consider
\be
\r^{ki} M = \r^k\r^i \left(-\r^i\r^j\r^k\right) = \r^k\r^j\r^k =
-(\r^k)^2\r^j=\r^j,
\ee
and we have thus proven cyclical invariance.  We then find that $T_2$
generalizes to
\be
T_2 = -i\int (\bar \l _A\left(\r^i\r^j\r^k\right)_{AB}\e_B)^n
\left[f^{amn}f^{apl} + f^{apn}f^{alm} + f^{aln}f^{amp} \right]
(\phi^i)^p(\phi^j)^l(\phi^k)^m
\ee
which vanishes similarly; $T_1$ is produced in the same way as for the
${\cal N}=4$ case.


\end{document}